\documentclass[aps,prl,twocolumn]{revtex4}

\usepackage{graphicx}%
\usepackage{dcolumn}
\usepackage{amsmath}
\begin{document}


\title{Spin Gunn Effect}

\author{Yunong Qi, Zhi-Gang Yu$^\dagger$, and Michael E. Flatt\'e}
\email{michael-flatte@uiowa.edu}
\thanks{\\ $\dagger$ now at SRI International, 333 Ravenswood Avenue, Menlo Park, CA 94025}
\affiliation{Optical Science and Technology Center and Department of Physics and Astronomy,
University of Iowa, Iowa City, Iowa 52242}

\date{\today}

\begin{abstract}
We predict that the flow of unpolarized current in electron-doped GaAs and InP at room temperature is unstable at high electric fields to the dynamic formation of spin-polarized current pulses.  Spin-polarized current is spontaneously generated because the conductivity of a spin-polarized electron gas differs from that of an unpolarized electron gas, even in the absence of spin-orbit interaction.  Magnetic fields are not required for the generation of these spin-polarized current pulses, although they can help align the polarization of sequential pulses along the same axis.  
\end{abstract}

\maketitle

Spin-based semiconductor electronics (``spintronics'') requires the generation of spin polarization in nonmagnetic semiconductors\cite{spintronics-book,sawolfscience}. Approaches through the injection of spin-polarized electrons or holes from magnetic materials\cite{Fiederling,Ohno,Zhu,Jonker,Crowell}, and spin-filtered\cite{spinfilter,spinfilter2} or spin-selective\cite{Dyakonov,Levitov,Edelstein,Zhang,katopreprint} currents using the spin-orbit interaction have been considered. These all rely on an energetic difference between spin-up and spin-down carriers in the electronic structure of some part of a circuit. Here we identify an entirely different approach. 

We show that initially unpolarized electron current flow in semiconductors can be unstable towards the spontaneous formation of spin-polarized current pulses even without an applied magnetic field.  The mechanism we propose is related to the charge Gunn effect\cite{Gunndisc}, in which homogeneous charge current flow is unstable to the spontaneous formation of high-electric-field domains, inhomogeneous charge distributions, and current pulses.   We predict the ``spin Gunn effect'' can be seen at room temperature and pressure in GaAs and InP, and possibly\cite{GaN} GaN.   A dependence of the electron drift velocity on spin polarization --- originating from the Pauli exclusion principle --- drives the spin Gunn effect.  A room-temperature source of spin-polarized electrons from a nonmagnetic semiconductor would significantly advance the range of realizable spintronic devices. The high-frequency oscillatory nature of these pulses also suggests new possibilities, such as may emerge from matching the oscillation frequency of the spin-polarized source to the precession frequency of spins in another material.

The instability of the spin Gunn effect originates from an electron velocity ($v=\mu E$, where $\mu$ is the mobility and $E$ the electric field) that (1) depends on the local spin polarization of the electrons, and (2) differs for spin-up and spin-down electrons. The spin-dependent mobility
\begin{equation}
\mu_{\uparrow(\downarrow)}  =  \mu(n/2)\cdot\left(1+(-)\alpha P\right)\label{mudeff}
 \end{equation}
where
\begin{equation}
P=(n_\uparrow- n_\downarrow)/(n_\uparrow + n_\downarrow)
\end{equation}
 is the spin polarization in the semiconductor, and $\alpha$ depends on the dominant {\it spin-conserving} electron scattering mechanism.

\begin{figure}[h]
\includegraphics[width=8cm]{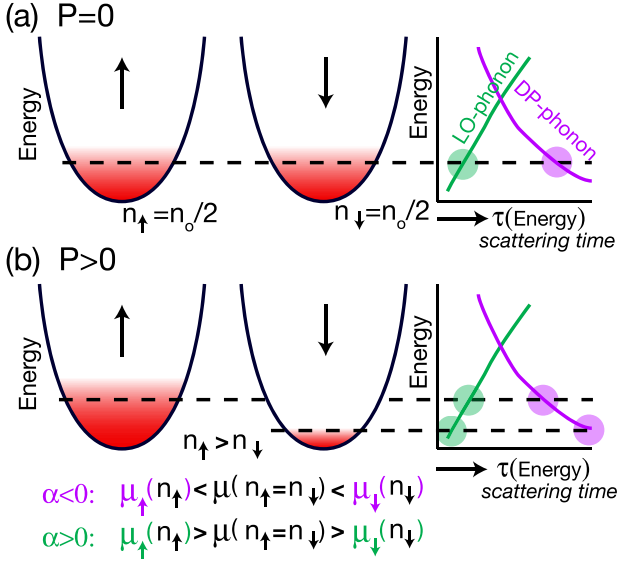}
\caption[]{The electron sea for spin-up and spin-down electrons in an electron-doped semiconductor. (a). The sea is not spin polarized, and the mobility of spin-up and spin-down electrons is identical. (b). The sea is spin polarized, and the mobility of spin-up electrons is either smaller (as in the case of DP-phonon scattering) or  larger (as in the case of LO-phonon, ionized-impurity, or piezoelectric-coupled-phonon scattering)  than the mobility of spin-down electrons.  In the first case $\alpha<0$ and in the second $\alpha>0$. }\label{mech}
\end{figure}

\begin{figure}[h]
\includegraphics[width=9cm]{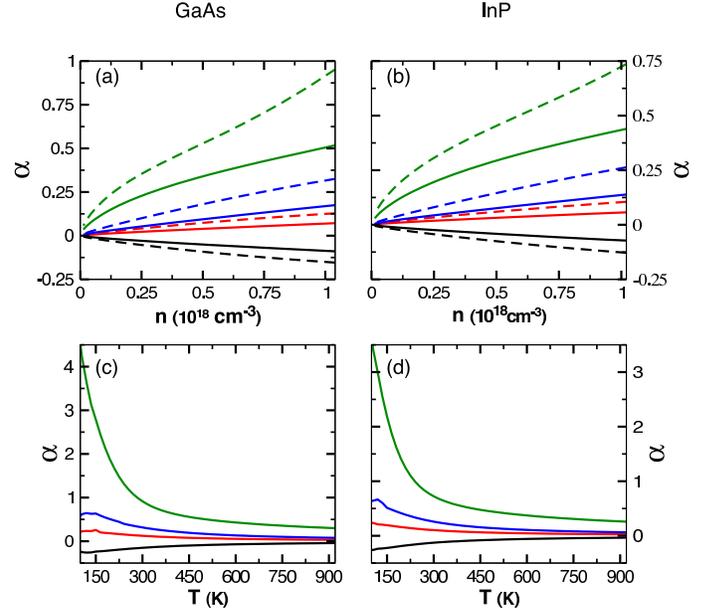}
\caption{(ab). $\alpha$ as a function of conduction electron density for GaAs and InP at 300K (dashed line) and 500K (solid line). (cd). $\alpha$ as a function of temperature for GaAs and InP for $n=10^{18}$~cm$^{-3}$ Green curves are LO-phonon scattering, blue curves are ionized impurity scattering, red curves are piezoelectric-phonon scattering, and black curves are DP-phonon scattering.}\label{alpha}
\end{figure}

Most semiconductors have $\alpha\ne 0$.  The common scattering processes for electrons are independent of the spin of the electron, but do depend on its energy\cite{Cardona-Yu}. As shown in Fig.~\ref{mech}, due to the Pauli exclusion principle the chemical potentials of spin-up and spin-down electrons in a spin-polarized electron gas differ from each other. As a result, for a degenerate electron gas the energy (and velocity) distribution of mobile spin-up electrons differs from that of spin-down electrons.  Therefore the mobility of electrons for $P\ne 0$ will differ from that when $P=0$.
For scattering of conduction electrons from ionized impurities, or acoustic phonons via piezoelectric coupling, or longitudinal phonons  via Fr\"ohlich coupling (LO-phonon scattering), the low-energy electrons are scattered more than the high-energy electrons\cite{Cardona-Yu}, as shown in Fig.~\ref{mech}. In the spin-polarized electron gas the increased energy of the spin-up electrons leads to longer scattering times and thus a higher mobility. The decreased energy of the spin-down electrons leads to shorter scattering times and a lower mobility. This situation corresponds to $\alpha>0$ in Eq.~(\ref{mudeff}). If the dominant scattering process involves acoustic phonons coupling via the deformation potential (DP-phonon scattering), then the spin-up electrons scatter more than the spin-down electrons\cite{Cardona-Yu}, as also shown in Fig.~\ref{mech}, and $\alpha<0$.  The only common scattering process that produces $\alpha = 0$ is neutral-impurity scattering. 
Thus the mobility of electrons in a spin-polarized semiconductor almost always depends strongly on the spin orientation, even when there is no explicit spin dependence in the scattering processes themselves.  At room temperature, and the elevated temperatures involved in the Gunn effect, LO-phonon scattering is likely to dominate, and we expect $\alpha >0$.  Calculations for GaAs and InP for $P\rightarrow 0$ of 
\begin{equation}
\alpha(P) = {n_\uparrow \mu_\uparrow - n_\downarrow\mu_\downarrow\over P(n_\uparrow\mu_\uparrow + n_\downarrow\mu_\downarrow)} - 1\label{alpha-P}
\end{equation}
 as a function of density at 300K and 500K, and as a function of temperature for $n=10^{18}$~cm$^{-3}$ are shown in Fig.~\ref{alpha}. The spin-polarization-dependence of the mobility increases with increasing density, and decreases with temperature, although $|\alpha|>0.1$ for all of these scattering processes for temperatures below 500K at $n=10^{18}$~cm$^{-3}$.

We now consider the effect of $\alpha \ne 0$  on the spin polarization of electrons in an initially unpolarized Gunn domain. 
The  Gunn domain itself moves in a self-sustaining way --- an inhomogeneous electric field moves the charge, and the shifting space-charge region moves the electric field  --- and the net result is the collective motion of the electric field shape and the space-charge region through the sample. In the presence of these electric fields a small difference in mobility between spin-up and spin-down electrons means fewer electrons of one spin species will flow in response to the field. This enhances the difference in mobility between spin-up and spin-down electrons, which leads to further differences in the accumulation or depletion of the two spin species. A positive feedback effect therefore drives the increase in the spin polarization.  The electric field is determined by the charge density through the Poisson equation, and does not depend on the spin polarization, so the domain continues to move as before but now with an inhomogeneous spin polarization.  We require only a small initial imbalance between spin-up and spin-down electrons to begin the effect. This can be nucleated thermally as an initial inhomogeneous spin polarization with a random orientation. It can also be nucleated with a small applied magnetic field, which will also serve to align the spin polarizations of all the domains. 

As the electron momentum scattering time is very short ($<100$~fs) we can calculate the time dependence of the spin polarization using spin-resolved drift-diffusion equations\cite{Fchapter},
\begin{eqnarray}
\frac{\partial n_\uparrow}{\partial t}&=&-\frac{n_\uparrow-n_\downarrow}{2T_1}-\frac{\partial\left(n_\uparrow v_\uparrow\right)}{\partial x}+D_\uparrow\frac{\partial^{2}n_\uparrow}{\partial x^{2}}\label{eq:8}\\
\frac{\partial n_\downarrow}{\partial t}&=&-\frac{n_\downarrow-n_\uparrow}{2T_1}-\frac{\partial\left(n_\downarrow v_\downarrow\right)}{\partial x}+D_\downarrow\frac{\partial^{2}n_\downarrow}{\partial x^{2}}\label{eq:9}\end{eqnarray}
where $T_1$ is the spin relaxation time of the electrons.   Using Eq.~(\ref{mudeff}), keeping only terms up to first order in $P$, neglecting pure diffusion effects and shifting to a moving reference frame with the domain, $x^{\prime}=x-v_{dom}t$ , where $v_{dom}$ is the velocity of the
domain,
\begin{eqnarray}
\frac{\partial P}{\partial t}  &=&  -\frac{\alpha }{n}\frac{\partial\left[nE\mu \right]}{\partial x^\prime}P - {P\over T_1} + (\mu(1-\alpha) E - v_{dom}){\partial P\over \partial x^\prime}\\
&\sim&(\gamma-{1\over T_1}) P 
\label{eq:19}\end{eqnarray}
The form of this equation yields spin amplification if $\gamma>T_1^{-1}$. The $\partial P/\partial x^\prime$ term describes the flow of inhomogeneous spin polarization from one region of the domain to another; similar terms appear in the charge Gunn effect itself and are ignored\cite{Kroemer-chapter} --- they change the quantitative results somewhat but not the qualitative behavior of the Gunn instability. In our treatment here we will neglect this term, which we calculate to be over an order of magnitude smaller than $\gamma P$.  The central quantity is the spin amplification rate $\gamma$.

The first observation about Eq.~(\ref{eq:19}) is that $\gamma$ is proportional to the spatial variation of the {\it drift} current. The drift current varies spatially due to the inhomogeneous electric fields and densities associated with the Gunn domain, and its derivative can be positive under certain conditions for $\alpha>0$ (e.g. downstream from the domain center) and under others for $\alpha<0$ (e.g. upstream). From the continuity equations Eqs.~(\ref{eq:8},\ref{eq:9}), $\partial n_s/\partial t  = \nabla\cdot {\bf j}_s$, so differing spin-up and spin-down currents (from different mobilities) produce different spin-up and spin-down density accumulation. Far away from the domain the local charge density is the equilibrium charge density, and $\gamma \rightarrow 0$.  To make $\gamma$ as large as possible we should consider cases involving large charge imbalances (within the breakdown fields of the material).  Ordinary doping levels for the charge Gunn effect are in the range of $10^{14}-10^{16}$~cm$^{-3}$. Doping levels of 10$^{18}$~cm$^{-3}$ lead to sharper domains and less microwave power in the fundamental mode of the oscillation, so they are not commonly used. For a spin Gunn effect, however, a sharp domain will lead to larger spin polarizations.

We now quantitatively calculate the value of $\gamma$ for a doping level of 10$^{18}$~cm$^{-3}$ at a 300K lattice temperature in GaAs and InP.   We ignore higher-order effects and evaluate  $-(\alpha/n)\partial [nE\mu(n/2)]/\partial x^\prime$  for $P=0$. Hence
\begin{equation}
\gamma  = -\alpha\mu E\left[\frac{e\left(n-n_{0}\right)}{\epsilon_{s}E}+\frac{1}{n}\frac{\partial n}{\partial x^{\prime}}+{1\over \mu}\frac{\partial \mu}{\partial x^{\prime}}\right].\label{eq:23}\end{equation}
The first term above comes from the derivative of the electric field, which is related to the local charge density through the Poisson equation. For $\gamma$ to be positive, if $\alpha<0$ the charge density must be larger than equilibrium ($n>n_o$), whereas if $\alpha>0$ the charge density must be depleted ($n<n_o$). Evaluation of $\gamma$ thus requires calculation of $n(x^\prime)$ and $E(x^\prime)$ for the Gunn domain.

\begin{figure}[h]
\includegraphics[width=9cm]{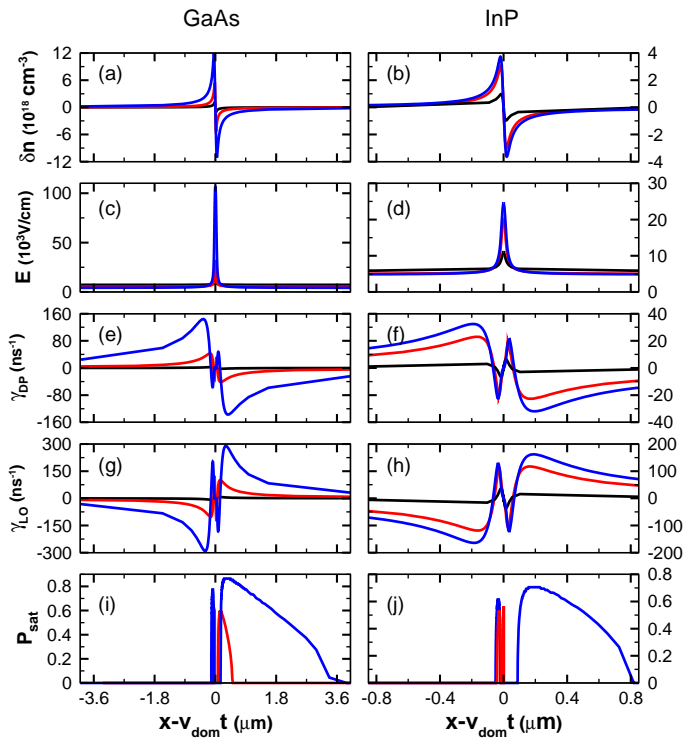}
\caption{The spatial dependence of the electron density $n$ [(a) for GaAs, (b) for InP] and the electric field $E$ [(c) for GaAs, (d) for InP] in the frame moving with the Gunn domain ($x^\prime=0$ is the center of the domain). The spin amplification rate $\gamma$ for DP-phonon scattering [(e) for GaAs, (f) for InP] and for LO-phonon scattering [(g) for GaAs, (h) for InP]. The three curves correspond to different drift velocities far from domain (different applied voltages to the Gunn diode): black is $5.7\times 10^6$~cm$/$s for GaAs, $2.6\times 10^7$~cm$/$s for InP, red is $5.0\times 10^6$~cm$/$s for GaAs, $2.45\times 10^7$~cm$/$s for InP, and blue is $4.0\times 10^6$~cm$/$s for GaAs, $2.4\times 10^7$~cm$/$s for InP. The saturation polarization (corresponding to $\gamma(P) = T_1^{-1}$) is shown in (i) for GaAs and (j) for InP.}\label{domain-dp}
\end{figure}

Our  calculation of $n(x^\prime)$ and $E(x^\prime)$  follows that of Sze\cite{Sze} for a mature, steady-state domain.  We assume the electrons in the lower valley and the upper valleys have the same temperature \cite{Hilsum,McCumber}, and that the
diffusion constant $D$ is independent of the electric field, the electron density and spin. The time-dependent electron current is equal to the displacement current,
\begin{equation}
J=env(E)-eD\frac{\partial n}{\partial x}= -\epsilon_{s}\frac{\partial E}{\partial t}.\label{timejE}\end{equation}
For a high-field domain propagating without a change of shape, in the moving frame ($x^{\prime}=x-v_{dom}t$) 
\begin{equation}
\frac{eD}{\epsilon_{s}}\frac{dn}{dE}=\frac{\left\{ n\left[v(E)-v_{dom}\right]-n_{0}\left(v_{R}-v_{dom}\right)\right\} }{n-n_{0}}.\label{dens-field}\end{equation}
The electron drift velocity outside of the domain, $v_R = J/(en_{0})$.
The solution of equation (\ref{dens-field}) is:
\begin{eqnarray}
 \frac{n}{n_{0}}-\ln(\frac{n}{n_{0}})-1&=&
 \frac{\epsilon_{s}}{en_{0}D}\int_{E_{R}}^{E}\Big\{ \left[v(E^{\prime})-v_{dom}\right]-\qquad
 \nonumber\\ &&\qquad\frac{n}{n_{0}}\left(v_{R}-v_{dom}\right)\Big\} dE^{\prime}
 \label{density}\end{eqnarray}
A self-consistent solution to Eq.~(\ref{density}) produces the required relationship between the density and the electric field $E(n)$. The spatial and temporal behavior of $E$ can then be found by evaluating Poisson's equation,
\begin{equation}
x^{\prime}=x_{0}^{\prime}+\frac{\epsilon_{s}}{e}\int_{E_{dom}}^{E}\frac{dE}{n-n_{0}}\label{poisson}\end{equation}
 where $x_{0}^{\prime}$ is the
location of the peak electric field in the moving frame of reference.  The local electron temperature $T_e$ is determined by the local electric field from
\begin{equation}
k_BT_e = k_BT_l + (2/3) q\tau_E \mu E^2 [1+R\exp(-\Delta E/k_B T_e)]^{-1}\label{temp-dep}
\end{equation}
where $T_l$ is the lattice temperature, $\tau_E$ is the energy relaxation time ($\sim 10^{-12}$s), $\Delta E$ is the energy separation between the $\Gamma$ and $L$ valleys, and $R$ is the ratio of the density of states in the $L$ valleys to the $\Gamma$ valleys. We have used values for these quantities from Sze\cite{Sze}.
Solutions of Eqs.~(\ref{density})-(\ref{poisson}) for the density and electric field as a function of position are shown in Fig.~\ref{domain-dp} for GaAs and InP for a 300K $T_l$. Fig.~\ref{domain-dp}(ab) show the density and Fig.~\ref{domain-dp}(cd) show the electric field.

The amplification factor $\gamma$ is shown as a function of position in Fig.~\ref{domain-dp}(ef) for the two materials if the dominant scattering process is DP-phonon scattering, and in Fig.~\ref{domain-dp}(gh) for LO-phonon scattering. As the electron temperature varies with position according to Eq.~(\ref{temp-dep}), $\alpha$ must be evaluated as a function of temperature and density to determine $\gamma$ properly. The electric field, and thus the spatial variation of the drift current, is largest at the center of the domain, however the electron temperature is also greatest there. These competing effects produce a maximum $\gamma$ which is near the domain but not at the peak of the space charge.  For GaAs there are regions of amplification for spin lifetimes as short as 3~ps, whereas for InP the amplification is present for lifetimes longer than 2~ps. The region of spin amplification differs for the two scattering mechanisms; for LO-phonon scattering the largest spin amplification is downstream of the domain, whereas for DP-phonon scattering the largest spin amplification is upstream of the domain.

The significant increase in $\gamma$ with doping density leads us to propose Gunn diodes with $n\sim 10^{18}$~cm$^{-3}$ for the spin Gunn effect. For a lower doping density of $10^{16}$~cm$^{-3}$ the $T_1$ of GaAs at 300K is 50~ps\cite{Optical-Orientation} and our calculated $\gamma\sim 4$~ns$^{-1}$, so spin amplification is not expected to occur. The elevated temperature near the Gunn regime, which ranges from 500K a distance $100$~nm away from the center of the domain, to much higher values near the peak electric field, will further reduce the spin relaxation times. Scaling with $T^3$ from D'yakonov-Perel' precessional relaxation\cite{Optical-Orientation} suggests spin relaxation times four times shorter at 500K, of the order of 12~ps. The reduced mobility at higher temperatures will lengthen these estimated times, and our calculated spin amplification rates of $\gamma(P=0) >0.4$~ps$^{-1}$ are five times larger than this rapid spin relaxation rate, providing confidence that spin amplification is likely for room-temperature devices.  The enhanced electron-electron scattering present in degenerate systems at these high temperatures is expected to further increase the spin relaxation times (without affecting the mobility)\cite{Ivchenko-deg}. Shown in Fig.~\ref{domain-dp}(ij) are steady-state (saturation) values of the spin polarization $(P_{sat})$ estimated for LO-phonon scattering determined by setting $T_1=10$~ps, using $\alpha(P\ne 0)$ in Eq.~(\ref{eq:23}), and solving $\gamma(P) = T_1^{-1}$. As $\alpha(P=1)=0$, $\alpha$ changes 100\% between $P=0$ and $P=1$. In contrast, the total $n$, total $\mu$, and $E$ change $\sim 20$\%. Thus we neglect nonlinear effects on the total $n$, total $\mu$, and $E$ in our calculation of $P_{sat}$. When $\gamma(P=0)<T_1^{-1}$ there is no amplification and $P_{sat}=0$. The largest $P_{sat}$ exceeds 80\% for both GaAs and InP, and should be directly visible in a Faraday rotation measurement (e.g. Ref.~\cite{Awschalom-GaN}). Therefore the spin Gunn effect is a pulse of highly spin polarized electrons located just before or just after a charge current pulse. 

The existence of this spin Gunn effect is exceptionally robust to temperature and spin relaxation, suggesting a wide range of potential applications of devices based on the effect. We have only considered here spin effects amenable to nearly analytic analysis, such as occur in the propagation of a ``mature domain''. Other known modes of operation of the charge Gunn effect, such as the limited space-charge accumulation (LSA) mode, will provide additional spintronic functionality when analyzed in the context of the instability to  non-zero spin polarization we have identified here. 

We acknowledge support from DARPA/ARO DAAD19-01-1-0490, and helpful discussions with D. R. Andersen, J. Levy, N. Samarth, and G. Vignale.


\bibliography{}




\end{document}